\documentclass[aps,prb,reprint,twocolumn,groupedaddress]{revtex4-1}
\usepackage{acronym,calc}
\usepackage{amssymb,amsmath,mathrsfs}
\usepackage{color,array,colortbl,multirow}
\usepackage[pdftex]{graphicx}
\usepackage{epstopdf}
\usepackage[bookmarks, 
plainpages=false]{hyperref}
\hypersetup{ 
    colorlinks,%
}
\usepackage{graphicx}
\usepackage{epstopdf}

\newcommand{\dif}[1]{\,\mathrm{d}#1}
\DeclareMathOperator{\RE}{\mathrm{Re}}
\DeclareMathOperator{\IM}{\mathrm{Im}}

\DeclareMathOperator{\ii}{i}
\newcommand{\eps}{\varepsilon_0}
\newcommand{\lo}[1]{_{\mathrm{#1}}}

\newcommand{\ket}[1]{|#1\rangle}
\newcommand{\unit}[1]{\,\,\mathrm{#1}}
\newcommand{\Ef}{\mathcal{E}}

\begin{document}
\title{A scheme comparison of Autler-Townes based slow light in inhomogeneously broadened quantum dot media}

\author{Per Lunnemann}
\email[]{plha@fotonik.dtu.dk}
\author{Jesper M\o rk}
\email[]{jesm@fotonik.dtu.dk}
\affiliation{Technical University of Denmark, \\Department of Photonics Engineering, 2800 Kgs Lyngby, Denmark}

\begin{abstract}
We propose a method to achieve significant optical signal delays exploiting the effect of Autler-Townes splitting in an inhomogeneously broadened quantum dot medium. The absorption and slow-down effects  are compared for three schemes i.e. $\Xi$, V and $\Lambda$, corresponding to different excitation configurations. Qualitative differences of the V-scheme compared to the $\Xi$- and $\Lambda$-scheme are found, which show that features of Autler-Townes splitting are only revealed in the V-scheme. The underlying physical mechanisms causing this discrepancy are analyzed and discussed. Finally we compare field propagation calculations of the schemes showing significantly larger achievable signal delays for the V-scheme despite finite absorption of the coupling field. This opens the possibility for using waveguide structures for both coupling and probe fields, thus significantly increasing the achievable signal delays.
\end{abstract}


\maketitle

\section{Introduction\label{introduction}}
A striking feature of a three level atom coherently driven by a strong coupling field and  probed by a weak probe field, is a resonance splitting of the  probed transition, known as \ac{ATS}. \cite{Autler1955} The dressing of the states leads to a strong change of the probe absorption and phase; the latter corresponds to a change of the group velocity. A special case of \ac{ATS} is \ac{EIT}, \cite{Fleischhauer2005} where the probe absorption vanishes within a narrow band due to destructive quantum interference among the different probe absorption paths.

By exploiting \ac{EIT} in a Na Bose-Einstein condensate, Hau et al. demonstrated in 1999 a tremendous reduction of the probe group velocity. \cite{Hau1999} Since then, astounding physical phenomena based on \ac{EIT} have been demonstrated, such as stored light, \cite{Liu2001,Phillips2001} stationary light,\cite{Bajcsy2003} single-photon slow-down and storage\cite{Eisaman2005} and  optical information processing with matter wave dynamics. \cite{Naomi2007} A device enabling significant control of the velocity of light is of interest for applications within optical communication and microwave photonics, see e.g. Ref. \onlinecite{Khurgin2009,Boyd2009} and references therein. Compact and cheap devices, however, are required for practical applications, which makes semiconductor material a desirable material for slow light devices.\cite{Mork2008}

Self-assembled \acp{QD}\cite{Michler2003} are excellent candidates for realizing \ac{EIT} in a solid material.\cite{Kim2004,Chang-Hasnain2003} \acp{QD} offer quantized energy levels with transition wavelengths that can be engineered. Furthermore, self-assembled \acp{QD} can be stacked to increase the \ac{QD} density and embedded in waveguide structures to increase the laser field overlap.\cite{Chang-Hasnain2003} However, there are several issues that hinder the achievement of \ac{EIT} in  \acp{QD}: Firstly, dephasing rates in \acp{QD} are significantly higher than those found in atomic gases due to coupling to phonon modes.\cite{Borri2001,Birkedal2001} While this problem may be dealt with by operating at cryogenic temperatures, current fabrication techniques are inevitably associated with an \ac{IHB} that, at cryogenic temperatures, is approximately three orders of magnitude larger than the homogeneous linewidth of the ground state transition of each individual \ac{QD}.  \ac{IHB} \cite{Heitz1999a,Heitz2001,Borri2006} is primarily associated with the size dispersion of the \acp{QD} \cite{Nakata1999} and tends to destroy the observable EIT features when operating on an ensemble. \cite{Kim2004,Lunnemann2009} As a result of these drawbacks, the associated slow light effects are typically mainly generated by the \ac{ATS} rather than the more delicate interference effect of \ac{EIT}.

Lately, \ac{ATS} in \acp{QD} have been reported using charged \acp{QD},\cite{Kroner2008} as well as neutral \acp{QD} based on level schemes exploiting either the \ac{FSS} \cite{Gammon1996,Marcinkevivcius2008,Xu2007} or the biexciton transition \cite{Jundt2008,Muller2008,Gerardot2009} (see Fig. \ref{fig:FSSschemes}). One advantage of using two non-degenerate fine structure levels in a V-type configuration, is the ease of separating the pump and probe due to the linear orthogonal polarizations. \cite{Xu2007} Furthermore, as will be shown later, such a configuration could potentially be used in an inhomogeneously broadened \ac{QD} ensemble. On the other hand, as pointed out in Ref. \onlinecite{Agarwal1997}, a V-type configuration does not offer the effect of \ac{EIT} since the two absorption paths of the dressed states always interfere constructively. Rather, the resulting reduced absorption and group velocity is solely a result of \ac{ATS}.
Furthermore, the specific scheme based on fine structure split transitions, requires both fields to be polarized in the plane of growth \cite{Seguin2005} (see Fig. \ref{fig:FSSschemes}). As a result, at least one of the two fields (coupling or probe) are required to propagate perpendicularly to the QD plane of growth. Letting the probe propagate perpendicularly to the plane, offers the possibility of using a waveguide structure for the coupling field, thereby reducing the required coupling power. However, this limits the interaction length of the probe field to only a few layers of \acp{QD}. Instead, if the waveguide structure is used for the probe, the coupling field must propagate perpendicular to the plane, and the beam profile needs to cover the full length of the waveguide. An unguided coupling field is impractical for integrated photonic devices, but equally important, the expanded mode area increases the power required for the coupling field. This problem is not present for the bi-exciton based scheme, where both fields are allowed to be either TE or TM polarized. However, as we shall see, this type of level configuration is particularly sensitive towards \ac{IHB}.\cite{Kim2004} Thus, to our knowledge, there are currently no experimental demonstrations of slow light based on \ac{ATS} in a semiconductor waveguide structure.

In this paper we propose a simple scheme that allows for \ac{ATS} based slow light in an inhomogeneously broadened \ac{QD} medium using a waveguide structure for both coupling and probe field. We start by introducing a simple but general level structure for which the response of the ensemble of \acp{QD} is analyzed. The calculated slow-down and absorption effects of the three schemes are compared and discussed. The dependence on decay time constants for the  V-scheme are further investigated in order to extract the underlying slow-down mechanisms. Finally, we present propagation simulations for the three schemes comparing the achievable signal delay for a given tolerated signal attenuation.

\section{Quantum dot model}
\begin{figure}
\begin{center}
\includegraphics[width=0.45\textwidth]{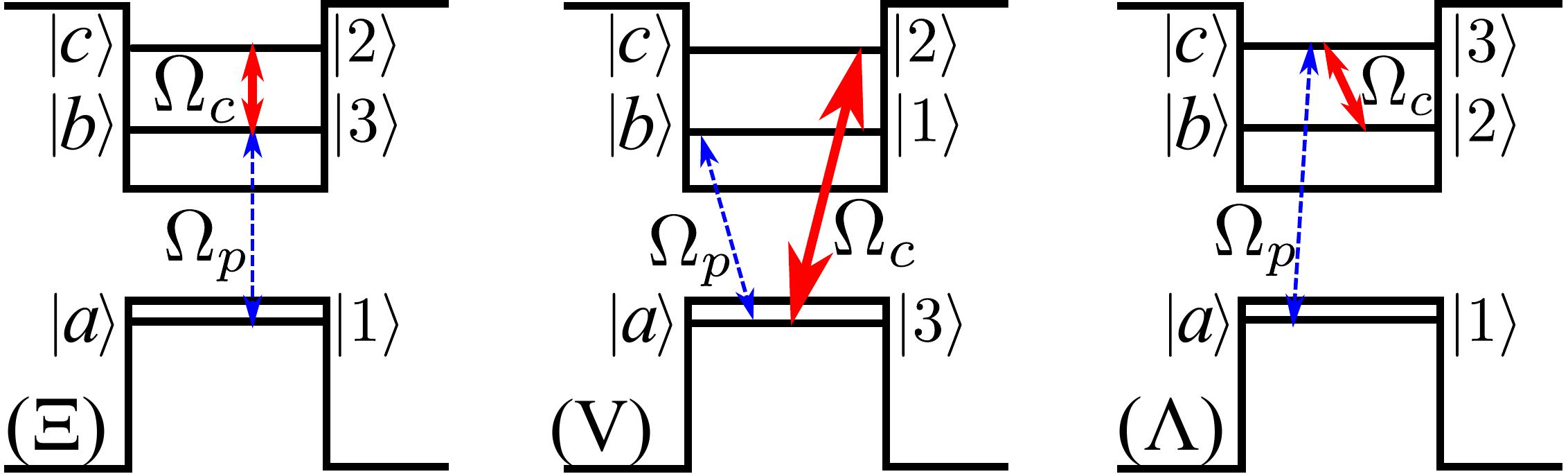} 
\caption{The three \ac{EIT} schemes $\Xi$ (left), V (middle) and $\Lambda$ (right) as implemented for a quantum dot confinement potential. Dashed blue and solid red arrows indicate the probe and coupling transition, respectively.\label{fig:schemes}}
\end{center}
\end{figure}

We consider \acp{QD} with discrete states, restricting our attention to two states in conduction band and a single state in the valence band. One transition is driven by a strong (coupling) laser field while a weak (probe) field probes a second transition. Three schemes $\Xi$, V and $\Lambda$ are considered that differ by the state used as a transit state, i.e. the state that is part of both the coupling and probe transition. The schemes are illustrated in Fig.  \ref{fig:schemes}.
For ease of notation and comparison among schemes, we shall denote the three states in two ways: $\ket{a}$, $\ket{b}$ and $\ket{c}$ refer to the valence band state and the two conduction band states, respectively, while $\ket{1}$, $\ket{2}$ and $\ket{3}$ are denoted in accordance with the considered transitions. I.e. the probe always couples the states $\ket{1}$ and $\ket{3}$ while coupling field couples $\ket{2}$ and $\ket{3}$.

Disregarding \ac{IHB} the following features are noted: For a symmetric confinement potential, the V- and $\Lambda$-scheme involve a forbidden transition. I.e. in a symmetric potential, the transition $\ket{1}\leftrightarrow\ket{3}$ is forbidden due to the orthogonality of the envelope wavefunctions. \cite{Coldren1997,Haug2004} Thus, in order to exploit the V- and $\Lambda$-scheme, one needs to break the symmetry of the QD potential. This can be achieved, e.g. by applying an electric DC field across the \ac{QD}. In reality, however, bandmixing and anisotropic strain effects relax these selection rules, thereby allowing for a finite coupling between the otherwise forbidden transitions, \cite{Stier1999,Barettin2009} however, with a somewhat weaker dipole moment. For the $\Lambda$ configuration this implies a reduction of the expected slow-down effect, while for the V-scheme a stronger coupling intensity is required for achieving the necessary \ac{ATS}. As mentioned, it is only the $\Xi$- and $\Lambda$-scheme that offer the possibility of \ac{EIT}. \cite{Agarwal1997} As such, for a single \ac{QD}, these schemes are expected to offer the most dramatic reduction of the group velocity and absorption. Especially the $\Xi$-scheme seems as the preferred choice, since the probe transition, as opposed to the $\Lambda$-configuration, possesses a large dipole moment. Because the group index roughly scales with the dipole moment squared,\cite{Kim2004} a large dipole moment is desirable for slow light effects.

\begin{figure}
\begin{center}
\includegraphics[width=0.45\textwidth]{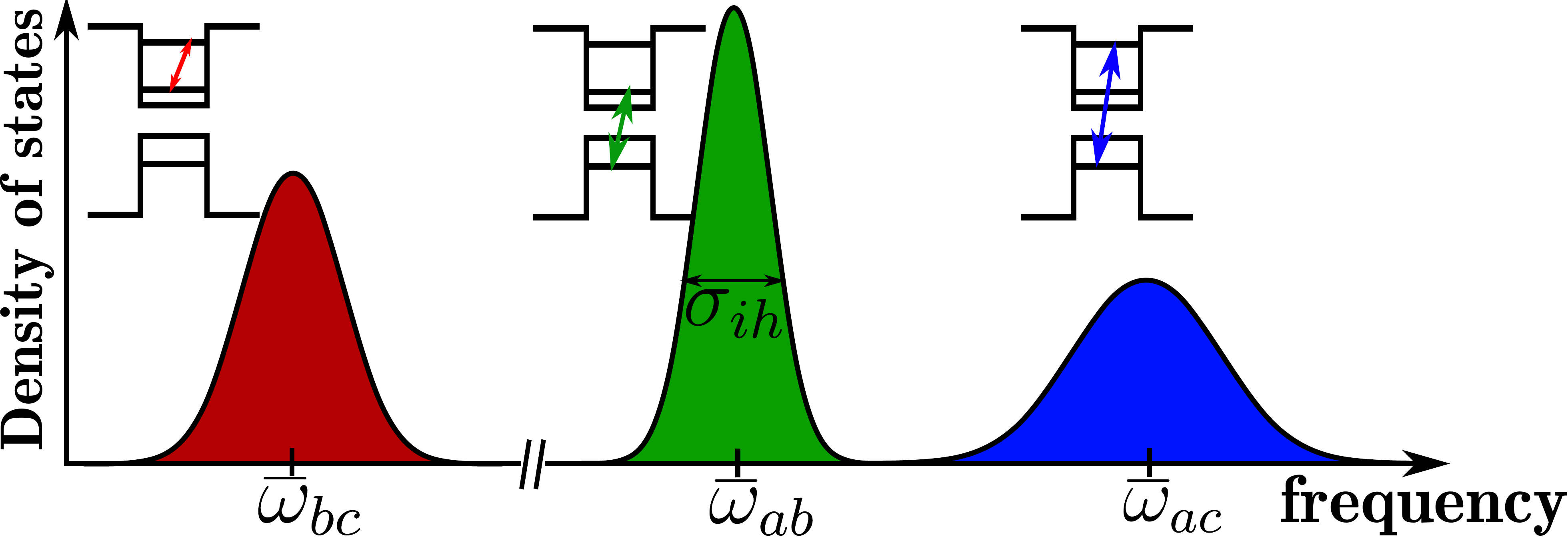} 
\caption{Density of resonant transitions as a function of frequency for the three considered transitions indicated by the level diagram insets. The IHB \ac{FWHM} is in this work referred to the ground state transition (middle peak). \label{fig:IHB}}
\end{center}
\end{figure}

The \ac{QD} is modeled as an infinite disc shaped confinement potential with radius $r$ and height $\varsigma$. In this case the undressed eigenenergies are found as:\cite{Kim2004}
\begin{equation}
\varepsilon_{nlm}(r,\varsigma)=\frac{\hbar^2}{2m_i^*}\left(\left(\frac{Z_{nl}}{r}\right)^2+\left(m\frac{\pi}{\varsigma}\right)^2  \right)\,,\label{eq:eigenenergy}
\end{equation} 
where $m_i$ is the effective electron or hole mass and $Z_{nl}$ denotes the $n$'th rooth of the $l$'th Bessel function, i.e. $J_l(Z_{nl})=0$.
Since the \ac{QD} sizes are dispersed in the ensemble, we explicitly write the radius, $r$, and height, $\varsigma$, dependence of the eigenenergy in Eq. \eqref{eq:eigenenergy}.
Assuming all \acp{QD} to have a constant aspect ration $\eta=\varsigma/r$, an infinitesimal volume change of the QD with a change of radius by $\delta r$ results in a shift of the eigenenergy of
\begin{equation}
\delta\varepsilon_{nlm}=-\frac{\hbar^2}{m_e^*}\frac{1}{r^3}\left[Z_{nl}^2+\left(\frac{m\pi}{\eta}\right)^2 \right]\delta r.
\end{equation}
Due to the heavier effective mass of the hole states, we consider only energy shifts of the electron states. Setting $\eta=3/10$ \cite{Heitz2001} the shift of the  first two electron states are found as
\begin{align}
\delta \varepsilon_{101}&\approx -\frac{\hbar^2}{r^3m_e^*}115.4\,\delta r \label{eq:firstexcited}\\
\delta \varepsilon_{111}&\approx -\frac{\hbar^2}{r^3m_e^*}62.17\,\delta r\,.\label{eq:secondexcited}
\end{align}
Linearizing the problem is justified since the inter-band transition energy is typically $\sim 1 \unit{eV}$, and the inhomogeneous broadening is on the order of a few tens of meV.\cite{Heitz2001} The shift of eigenenergies leads to a shift of the optical transitions that we shall refer to as a spectral shift and is illustrated in Fig. \ref{fig:IHB}.
For a given subensemble of QDs with fixed energies, we denote the spectral shift of the probe (coupling) transition $\Delta_{ih,p}$ $(\Delta_{ih,c})$ and define it relative to the mean probe (coupling) transition energy $\bar{\omega}_{13}$ of the QD ensemble, i.e. $\Delta_{ih,p}\equiv\omega_{13}-\bar{\omega}_{13}$ ($\Delta_{ih,c}\equiv\omega_{23}-\bar{\omega}_{23}$).
Relating the spectral shift of the probe transition, $\Delta_{ih,p}$, to the spectral shift of the coupling transitions $\Delta_{ih,c}$ we find for the three schemes:
\begin{subequations}\label{eq:kappa}
\begin{eqnarray}
\Xi\textrm{: }\Delta_{ih,c}=&\frac{\delta (\varepsilon_{111}-\varepsilon_{101})}{\delta \varepsilon_{101}}\Delta_{ih,p}&  \equiv\kappa\Delta_{ih,p}\approx 0.077\Delta_{ih,p}\\
\textrm{V: }\Delta_{ih,c}=&\frac{\delta \varepsilon_{111}}{\delta \varepsilon_{101}}\Delta_{ih,p}&\equiv\kappa\Delta_{ih,p}\approx 1.08\Delta_{ih,p}\label{eq:shiftV}\\
\Lambda\textrm{: }\Delta_{ih,c}=&\frac{\delta \left(\varepsilon_{111}-\varepsilon_{101}\right) }{\delta \varepsilon_{111}}\Delta_{ih,p}&\equiv\kappa\Delta_{ih,p}\approx 0.072\Delta_{ih,p},
\end{eqnarray}
\end{subequations}
where $\kappa\equiv\Delta_{ih,c}/\Delta_{ih,p}$. 

The absorption coefficient and group index are related to the imaginary and real part of the electric susceptibility $\chi_p$ associated with the probe transition. $\chi_p$ is calculated from the density matrix equations assuming monochromatic classical electric probe and coupling fields  and phenomenologically adding population and polarization decay rates.\cite{Fleischhauer2005} For one sub-ensemble of QDs with fixed energy levels (size), $\chi_p$ calculated to first order in the probe field evaluates to:
\begin{subequations}
\begin{widetext}
\begin{equation}
\mathrm{V}: \chi_{p}(\widetilde{\Delta}_{p},\widetilde{\Delta}_{c})=
\frac{\Gamma}{V}\frac{\mu_{13}^2}{\eps\hbar}2\frac{\{\gamma_{23}\left[\ii\Gamma_{13}\Gamma_2-2\Gamma_1\left(\Delta_p+\ii\gamma_{12}\right)\right]+\Delta_c(\Gamma_{13}\Gamma_2+2\gamma_{23}\Gamma_1)\}\Omega_c^2-2\left(\delta_-+\ii\gamma_{12}\right)\zeta}{\left[4 \left(\delta_{-} +i \gamma _{12}\right) \left(i \gamma _{13}+\widetilde{\Delta} _p\right)-\Omega _c^2\right] \left[ \zeta +\gamma _{23} \left(2 \Gamma _{13}+\Gamma _{12}\right) \Omega _c^2\right]}\label{eq:chiV}
\end{equation}
\end{widetext}
\begin{align}
\boldsymbol{\Xi}: \chi_{p}(\widetilde{\Delta}_{p},\widetilde{\Delta}_{c})&=\frac{\Gamma}{V}\frac{\mu_{13}^2}{\eps\hbar} 
\frac{2 (\delta_{+}+i \gamma_{12})}{\Omega_{c}^2-4 \left(\delta_{+}+i \gamma_{12}\right) (\widetilde{\Delta}_{p}+i \gamma_{13})}\label{eq:chiC}\\
\boldsymbol{\Lambda}: \chi_{p}(\widetilde{\Delta}_{p},\widetilde{\Delta}_{c})&=\frac{\Gamma}{V}\frac{\mu_{13}^2}{\eps\hbar}
\frac{2\delta_{-}+i \gamma_{12} }{\Omega_c^2-4 (\delta_{-}+i \gamma_{12}) (\widetilde{\Delta}_{p}+i \gamma_{13})},\label{eq:chiL}
\end{align}
\end{subequations}
where $V$ is the QD volume, $\Gamma$ is the confinement factor and $\varepsilon_0$ is the vacuum permittivity. $\widetilde{\Delta}_{p}$ and $\widetilde{\Delta}_{c}$ are the effective probe and coupling laser detuning, i.e. $\widetilde{\Delta}_p=\Delta_p-\Delta_{ih,p}$ and $\widetilde{\Delta}_c=\Delta_c-\kappa\Delta_{ih,p}$ where $\Delta_{p}=\omega_{p}-|\omega_{3}-\omega_{1}|$ and $\Delta_{c}=\omega_{c}-|\omega_{3}-\omega_{2}|$. A two-photon detuning is defined as $\delta_{\pm}\equiv\widetilde{\Delta}_{p}\pm\widetilde{\Delta}_{c}$, $\mu_{13}$ is the dipole moment of the probe transition and the Rabi frequency, $\Omega_c$ of the coupling field is defined as $\Omega_c \equiv \boldsymbol{\mu_{23}}\cdot\mathcal{E}_c/(2\hbar)$ where $\mathcal{E}_c$ is the coupling electric field amplitude. Population and dephasing (polarization) decay rates between the states $i$ and $j$  are denoted $\Gamma_{ij}$ and $\gamma_{ij}$, respectively. In the expression for the V scheme we furthermore defined $\Gamma_{1}=\Gamma_{13}-\Gamma_{12}$ and $\Gamma_{2}=\Gamma_{23}+\Gamma_{12}$, i.e. the total population loss rate of state 1 and 2, respectively. Finally, we defined $\zeta=2\Gamma_{13}\Gamma_{2} (\widetilde{\Delta}_{c}^2 + \gamma_{23}^2)$. Comparing equations \eqref{eq:chiC}, \eqref{eq:chiL} and \eqref{eq:chiV}, it is seen that the susceptibility only depends on the homogeneous linewidths for the $\Xi$- and $\Lambda$-scheme, whereas for the V-scheme  the population decay rates also become important. As we shall see later, this puts restrictions on the population decay rates in order to achieve \ac{ATS} using the V-scheme.
Denoting the inhomogeneous distribution function of the probe transition by $f$, the mean electric susceptibility is obtained by averaging over the QD ensemble:
\begin{equation}
\langle\chi(\Delta_p)\rangle=
\int\limits_{-\infty}^{\infty}f(\Delta_{ih})\chi(\Delta_p-\Delta_{ih},\Delta_c-\kappa\Delta_{ih})\,\mathrm{d}\Delta_{ih},\label{eq:avchi}
\end{equation}
where the subindex $p$ was omitted in $\Delta_{ih}$.

To the author's knowledge, no publications are available, giving the measured population decay and dephasing rates on the same QD sample for transitions related to those depicted in Fig. \ref{fig:schemes}. Thus, the population decay rates are set similar to those reported by Heitz et al..\cite{Heitz2001} We note though, that in this paper they report on \acp{QD} with an exceptionally low intra-band population decay rate. It is demonstrated, that the geometrical shape of the \acp{QD} can be tailored to either enhance \cite{Heitz1999} or reduce \cite{Heitz2001} the exciton-LO-phonon coupling. In the latter case, truncated pyramidal shaped \acp{QD} where used, giving measured intra-band relaxation rates $\sim$15 times lower than the radiative population decay rate of the ground state transition. Later, we shall discuss the consequences of using typical \acp{QD} with a much faster intra band decay rate. For the $\ket{c}\rightarrow\ket{a}$ decay, we tentatively set $\Gamma_{ac}=\Gamma_{ab}$. We note, though, that for the considered case of a closed 3-level system, the transition rate $\Gamma_{ac}$ is governed by the dipole moment squared, cf. Fermi\rq{}s golden rule. \cite{Bransden2003} Since $\mu_{ac}$ is set very low, a lower value of $\Gamma_{ac}$ relative to $\Gamma_{ab}$ is predicted by Fermi\rq{}s golden rule. In reality however, carrier-carrier scattering and carrier-phonon interaction\cite{Markussen2006} lead to larger decay rates,  i.e. the system is not truly a closed 3-level system.

Concerning the dephasing rates, earlier measurements show near lifetime limited decay rates of the ground state transition near $0 \unit{K}$ temperature.\cite{Borri2001,Langbein2004,Borri1999a} Thus, for the calculations we tentatively set the dephasing rates of all transitions to be lifetime limited unless otherwise stated. The parameter values are given in table \ref{tab:params}.
\begin{table}
\begin{center}
\begin{tabular}{crrr}
\hline\hline
\textbf{Parameter} & $\ket{a}\leftrightarrow\ket{b}$ & $\ket{b}\leftrightarrow\ket{c}$ & $\ket{a}\leftrightarrow\ket{c}$\\\hline
$\eta$ & \multicolumn{3}{c}{$3/10$} \\
$\Gamma$ & \multicolumn{3}{c}{$6\cdot 10^{-3}$} \\
$V$ & \multicolumn{3}{c}{$1200\unit{nm^3}$} \\
$\sigma_{ih}$ & \multicolumn{3}{c}{$10\unit{meV}$} \\
$\hbar\omega_{ij}$ & $0.999\unit{eV}$ & $61\unit{meV}$& $1.06\unit{eV} $ \\
$\mu_{ij}$ & 0.7 $e\cdot$nm  & 4.7 $e\cdot$nm & 0.10 $e\cdot$nm \\
$\Gamma_{ij}$ & $2.6\unit{\mu eV}$ & $0.16 \unit{\mu eV}$  & $2.6\unit{\mu eV}$\\
\hline\hline
\end{tabular}
\end{center}
\caption{Quantum dot parameters used in the calculations unless otherwise stated. $\eta$ is the average height-radius ratio, $\Gamma$ is the confinement factor, $V$ is the average QD volume and $\sigma_{ih}$ is the \ac{FWHM} of the spectral distribution function $f$. $\hbar\omega_{ij}$, $\mu_{ij}$ and $\Gamma_{ij}$ are the energy difference, the dipole moment and the population decay rate of the transition between $\ket{i}$ and $\ket{i}$. The homogenous linewidths $\gamma_{ij}$ are all assumed lifetime limited.}
\label{tab:params}
\end{table}

\section{Slowdown and Absorption calculations}
\subsection{Scheme comparison}\label{sec:schemeComp}
\begin{figure}
\begin{center}
\includegraphics[width=0.45\textwidth]{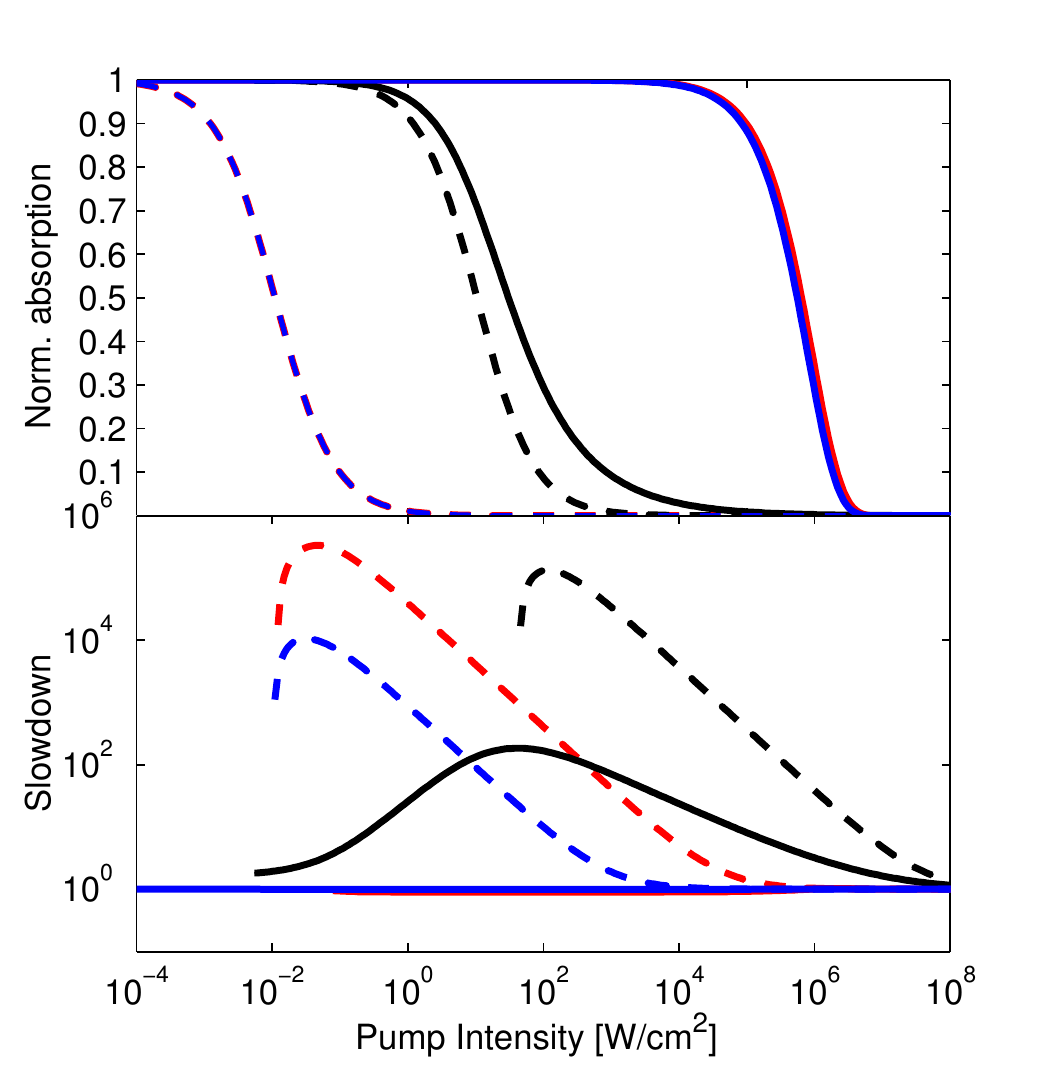} 
\caption{(color online) Top: Normalized absorption as  a function of coupling intensity for the three schemes $\Xi$ (red), V (black) and $\Lambda$ (blue), with (solid) and without (dashed) \ac{IHB}. The absorption is normalized by the absorption without an applied coupling field. Note that the curves for $\Xi$ and $\Lambda$ overlap. Bottom: corresponding slowdown factor $S$. Material parameters are presented in table \ref{tab:params}.}
\label{fig:slowdownComp}
\end{center}
\end{figure}

As a figure of merit describing the degree of group velocity control, we define the slow-down factor as $S=n_{g}/n_{bg}$,
where $n_{bg}$ is the refractive  index of the surrounding lossless background material while $n_{g}$ is the calculated group velocity including the \ac{QD} material and given as
\begin{equation}
n_g=\RE\left(\sqrt{1+\chi_{bg}+\chi_p}\right)+\frac{\omega\partial\left(\RE[\sqrt{1+\chi_{bg}+\chi_p}]\right)}{\partial\omega},
\end{equation}
where $\chi_{bg}$ is the electric susceptibility of the background material.
In Fig. \ref{fig:slowdownComp} the probe absorption (top) normalized by the absorption without a coupling field and the probe slow-down factor (bottom) is seen plotted as a function of coupling intensity for the three different schemes using \eqref{eq:chiC}- \eqref{eq:chiV}. Both probe and coupling detuning are set on resonance with the average transition frequencies $\bar{\omega}_{13}$ and $\bar{\omega}_{23}$, respectively.
Calculations with (solid) and without (dashed) \ac{IHB} are presented.
Comparing the three schemes without \ac{IHB}, the maximum slow-down of the V-scheme requires a stronger coupling intensity compared to the other two scheme. This is a result of the weak dipole moment between $\ket{a}$ and $\ket{c}$. Thus, for the V-scheme, a fairly high intensity is required in order to realize the required \ac{ATS}. The $\Lambda$-scheme, shows a somewhat lower slow-down factor as a result of the low dipole moment of its probe transition. From the absorption plot, we note that the $\Xi$- and $\Lambda$-schemes overlap due to the identical dipole moment of the coupling transitions and the similar dephasing rates $\gamma_{12}$ and $\gamma_{13}$, see equation \eqref{eq:chiC} and \eqref{eq:chiL}.
When comparing the calculated slow-down factors including \ac{IHB}, it is seen that the V-scheme is far superior. For the two other schemes, the inhomogeneous broadening is seen to completely destroy any slow-down effect, whereas  the V scheme maintains a slow-down factor of more than  $10^2$. Also, the required coupling  power for optimal slow-down is smaller by roughly 3 orders of magnitude, relative to the other two schemes. Finally, the V-scheme shows an almost similar absorption as a function of coupling intensity when including \ac{IHB}. This certainly is not the case for the $\Xi$- and $\Lambda$-scheme, where the coupling field needs to be increased by more than 8 orders of magnitude to achieve the same reduction of absorption.

\begin{figure}
\begin{center}
\includegraphics[width=0.46\textwidth]{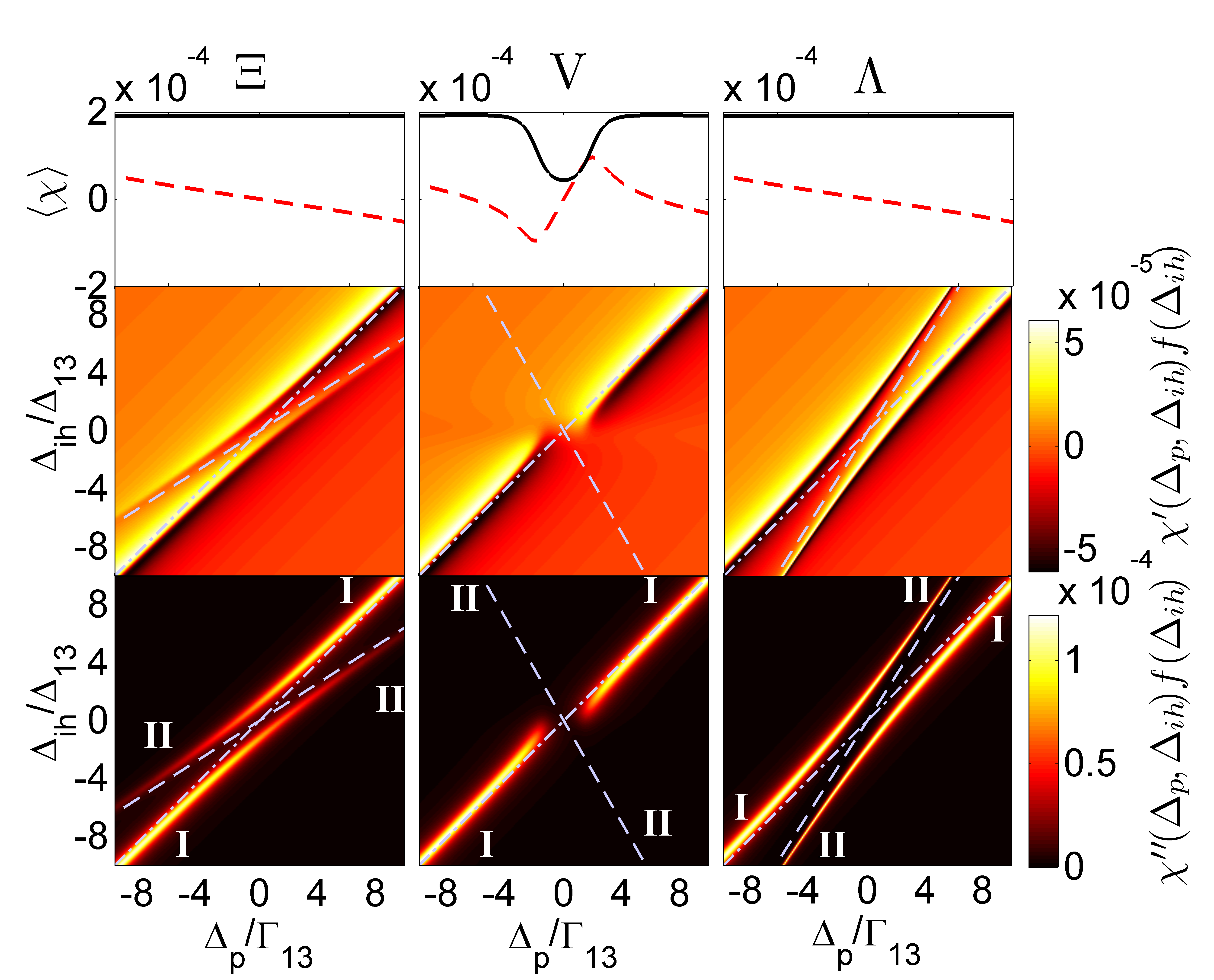} 
\caption{Electric susceptibility (in units of $\Gamma\mu_{13}^2/(V\eps\hbar$)) for QDs with $\eta=1$.  Left middle and right column is for the $\Xi$, V and $\Lambda$, respectively. Top: Average real (dashed) and imaginary (solid) part of the susceptibility as a function of the normalized probe detuning. Middle: Real part of the electric susceptibility as a function of the	 normalized probe detuning, $\Delta_{p}$, and spectral shift $\Delta_{ih}$. Bottom: Imaginary part of the electric susceptibility as a function of the normalized probe detuning, $\Delta_{p}$, and spectral shift $\Delta_{ih}$. The asymptote of the primary and secondary resonance is indicated by a dash-dotted and dashed line, respectively.}
\label{fig:chiComparison}
\end{center}
\end{figure}

To understand these notable differences we need to discuss the details of the \ac{ATS} mechanism. The presence of a strong coupling laser turns the bare eigenstates $\ket{2}$ and $\ket{3}$ into the dressed states $\ket{+}$ and $\ket{-}$ that appear as two absorption resonances for the probe. In a rotating frame the interaction Hamiltonian can be written:\cite{Fleischhauer2005}
\begin{equation}
\hat{H}=-\hbar
\begin{pmatrix}
\Delta_1	&	0	 	& 	\Omega_p\\
0 		&	\Delta_2 	 & 	\Omega_c\\
\Omega_p 	& 	\Omega_c	&	0
\end{pmatrix}
\end{equation}
where $\Delta_1=\Delta_p-\Delta_{ih}\,\,(\Delta_1=-\Delta_p+\Delta_{ih})$ for the V ($\Xi$ and $\Lambda$) scheme and $\Delta_2=\Delta_c-\kappa \Delta_{ih}\,\, (\Delta_2=-\Delta_c+\kappa \Delta_{ih})$ for the V and $\Xi$ ($\Lambda$) scheme. Letting $\Omega_p\rightarrow 0$ we find the eigenvalues as
\begin{equation}
\lambda_1=-\Delta_1,\quad\lambda_{\pm}=\frac{1}{2}\left(-\Delta_2\pm\sqrt{4\Omega_c^2+\Delta_2^2}\right)\label{eq:eigenvalues}
\end{equation}
where the $\lambda_\pm$ are the shifts of the dressed resonances of $\ket{3}$. Thus, the resonance condition for the probe transition is shifted accordingly. For a class of \acp{QD} spectrally shifted by $\Delta_{ih}$, this becomes
\begin{equation}
\Xi,\Lambda\textrm{: } \Delta_p=\Delta_{ih}+\lambda_\pm\qquad \textrm{V: } \Delta_p=\Delta_{ih}-\lambda_\pm\label{eq:rescond}
\end{equation}
Suppose the probe laser is on resonance with QDs at the center of the distribution function $f_{ih}$, i.e. $\Delta_{p}=\Delta_{ih}=0$. Since the coupling field has shifted the resonance by $\lambda_\pm$, QDs at the center of the distribution function do not contribute with absorption,assuming the coupling Rabi frequency larger than the linewidth of the probe transitions. However, for $\Delta_{ih}\neq 0$ this may not be the case. Rewriting \eqref{eq:rescond} for $\Delta_p=0$ we find:
\begin{equation}
\left([2\pm\kappa]^2-\kappa^2\right)\Delta_{ih}^2=4\Omega_c^2,\label{eq:rescond2}
\end{equation}
where the sum is for the $\Xi$-scheme and the difference is for the V- and $\Lambda$-scheme. A real solution to \eqref{eq:rescond2} for $\Delta_{ih}$ exists when
\begin{equation}
\mathrm{\Xi:}\,\,\kappa>-1\qquad\textrm{and }\qquad\mathrm{V,\Lambda:}\quad\kappa<1\label{eq:kappacond}
\end{equation} 
Comparing with the calculated values of $\kappa$ in \eqref{eq:kappa}, using the discpotential, we see that \eqref{eq:kappacond} is only fulfilled for the $\Xi$- and $\Lambda$-scheme. Physically, this means that for the V-scheme, none of the QDs in the inhomogeneously broadened ensemble have one of the two dressed states on resonance with the probe, as opposed to the other two schemes. Importantly, this implies that the probe in the V-scheme does not experience significant absorption from any of the QDs in the ensemble. To this we may add that we expect this observation holds true even when using a more realistic QD model, e.g. by using a more realistic QD potential as well as including the energy  shifts of the holes. Generally, one finds that $\kappa>1$, $\kappa>1$ and $0<\kappa<1$ for the $\Xi$-, V- and $\Lambda$-scheme respectively. This difference is clearly seen in Fig. \ref{fig:chiComparison} where the real and imaginary parts of the electric susceptibility $\chi_p$ are plotted as a function of $\Delta_p$ and $\Delta_{ih}$.
Left, middle, and right column are the electric susceptibility of the $\Xi$-, V- and $\Lambda$-schemes, respectively, with the top and bottom plots showing the real and imaginary part of the susceptibility $\chi_{p}$ normalized by $\left[\eps\hbar V(\Gamma\mu^{2})^{-1}\right]$. Without loss of generality, we have in this plot, for the purpose of illustration, used a height-radius aspect ratio $\eta=1$, although this ratio is typically found smaller.\cite{Heitz2001} The plot clearly shows two resonances crossing the region where $\Delta_{p}=0$ for the $\Lambda$ and $\Xi$ scheme. Thus, the average susceptibility, plotted in the top row of Fig. \ref{fig:chiComparison} using equation \eqref{eq:avchi}, reveals that the transparency region is destroyed and slow light effects vanish.

An illustrative way of exposing the difference between the schemes is by considering the resonance condition \eqref{eq:rescond} for \acp{QD} largely detuned from the center of $f_{ih}$, i.e. $|\Delta_{ih}|\gg\Omega_c$. Using the limit of \eqref{eq:eigenvalues} for $|\Delta_{2}|\rightarrow\infty$, we find the two approximate resonance conditions:
\begin{equation}
\textrm{I: }\Delta_{ih}=\Delta_{p}\quad\textrm{and}\quad\textrm{II: }\Delta_{ih}=\left(1\pm\kappa\right)^{-1}\Delta_{p},\label{eq:limres}
\end{equation}
where the sum is for the $\Xi$-scheme while the difference is for V- and $\Lambda$-schemes. We denote the two resonances the \emph{primary} and \emph{secondary} resonance.\cite{Lunnemann2009} The two asymptotes are illustrated as a solid and dashed line in Fig. \ref{fig:chiComparison}. By virtue of the values of $\kappa$ in \eqref{eq:kappa}, the asymptotes of $\Delta_{ih}$ vs. $\Delta_p$ have a positive slope for the $\Xi$- and $\Lambda$-scheme. From Fig. \ref{fig:chiComparison}, this clearly implies that for these schemes, the resonance crosses the region where $\Delta_{p}=0$.  The V-scheme, however, exhibits a negative slope, and as a result, no dressed state is resonant with the probe field.

Another notable difference between the V-scheme compared to the two other is that only near $\Delta_{ih}\approx 0$ do two resonances occur, i.e. the secondary resonance vanishes. In the bare state picture, the primary resonance is interpreted as the absorption from $\ket{1}\rightarrow\ket{3}$ while the secondary resonance is, for $\Xi$ and $\Lambda$, the sum of absorption pathways $\ket{1}\!\rightarrow\left(\!\ket{3}\!\rightarrow\!\ket{2}\!\rightarrow\!\ket{3}\right)^n$ where n is the number of Rabi floppings on the coupling transition. For the V-scheme the  pathways are  $\left(\ket{3}\!\rightarrow\!\ket{2}\!\rightarrow\!\ket{3}\right)^n\!\rightarrow\!\ket{1}$, where the coupling field is seen to excite carriers. Thus, for those \acp{QD} where the coupling laser is far off resonance, i.e. $|\Delta_{ih}|\gg0$, carriers from $\ket{3}$ are no longer excited to $\ket{2}$. As a result, the absorption pathways associated with the secondary resonance vanishes in the V-scheme.

\subsection{V-scheme parameter dependence}
As mentioned, we have chosen a very low intraband decay rate  $\sim 8\unit{ns}$. Although such QDs have been fabricated,\cite{Zibik2009,Heitz2001} quantum dots typically show intraband relaxation on a time scale of $\sim10\unit{ps}$.\cite{Heitz1999a} 
\begin{figure}
\begin{center}
\includegraphics[width=0.45\textwidth]{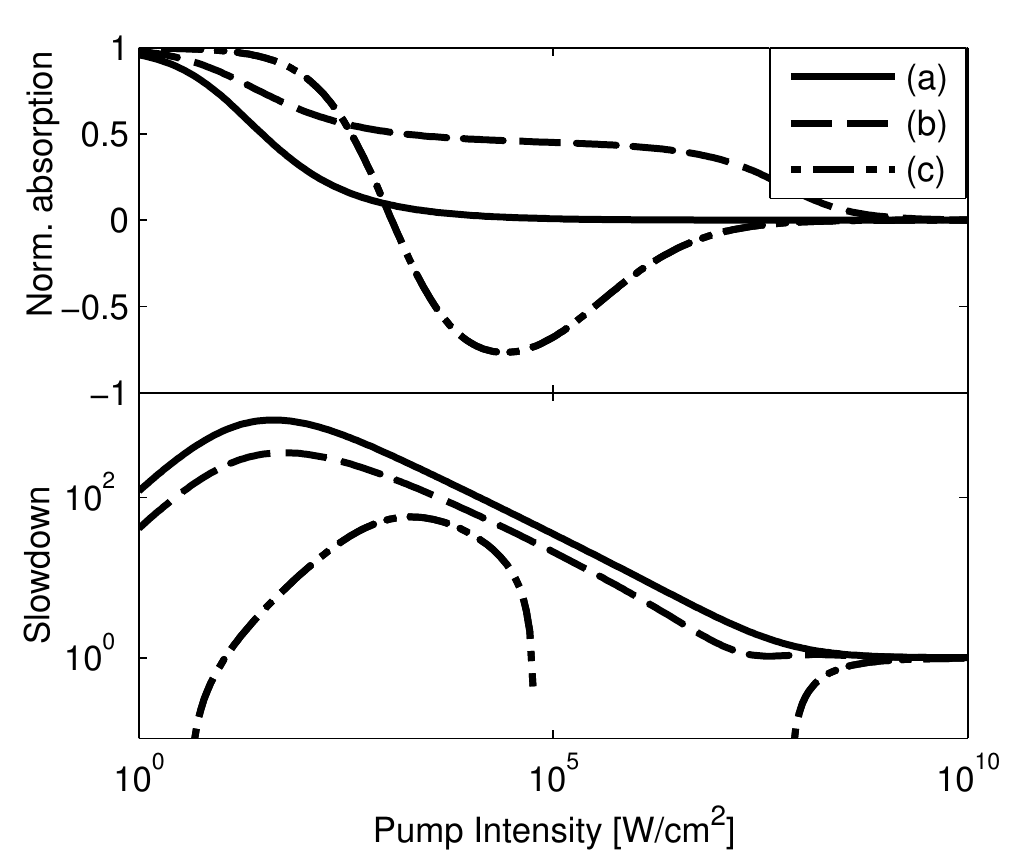} 
\caption{Top: absorption as a function of coupling field intensity for the three cases: (a, solid) lifetime limited dephasing rates with no intraband population decay rate, (b, dashed) lifetime limited dephasing rates with a large intraband population decay rate and (c, dotted) being equivalent to (a) except the intraband dephasing is set large. Bottom: corresponding slow-down factors. \label{fig:paramInvestigation}}
\end{center}
\end{figure}
In Fig. \ref{fig:paramInvestigation} we compare the absorption and slow-down of the V-scheme in three cases corresponding to different decay time constants: 
\begin{itemize}
\item[(a)] Lifetime limited polarization decay with low  intraband population decay rate; $\Gamma_{12}/\Gamma_{13}=0$.
\item[(b)] Rates equivalent to a) except the intraband polarization decay $\gamma_{12}$ is set a thousand times larger than the lifetime limited value, i.e. $\gamma_{12}=1000\cdot\frac{1}{2}\left(\Gamma_{13}+\Gamma_{23}+\Gamma_{12}\right)$.
\item[(c)] Lifetime limited polarization decay and a large intraband population decay rate;  $\Gamma_{12}/\Gamma_{13}=25$, corresponding to a relaxation time of 10 ps. 
\end{itemize}
\begin{figure}
\begin{center}
\includegraphics[width=0.38\textwidth]{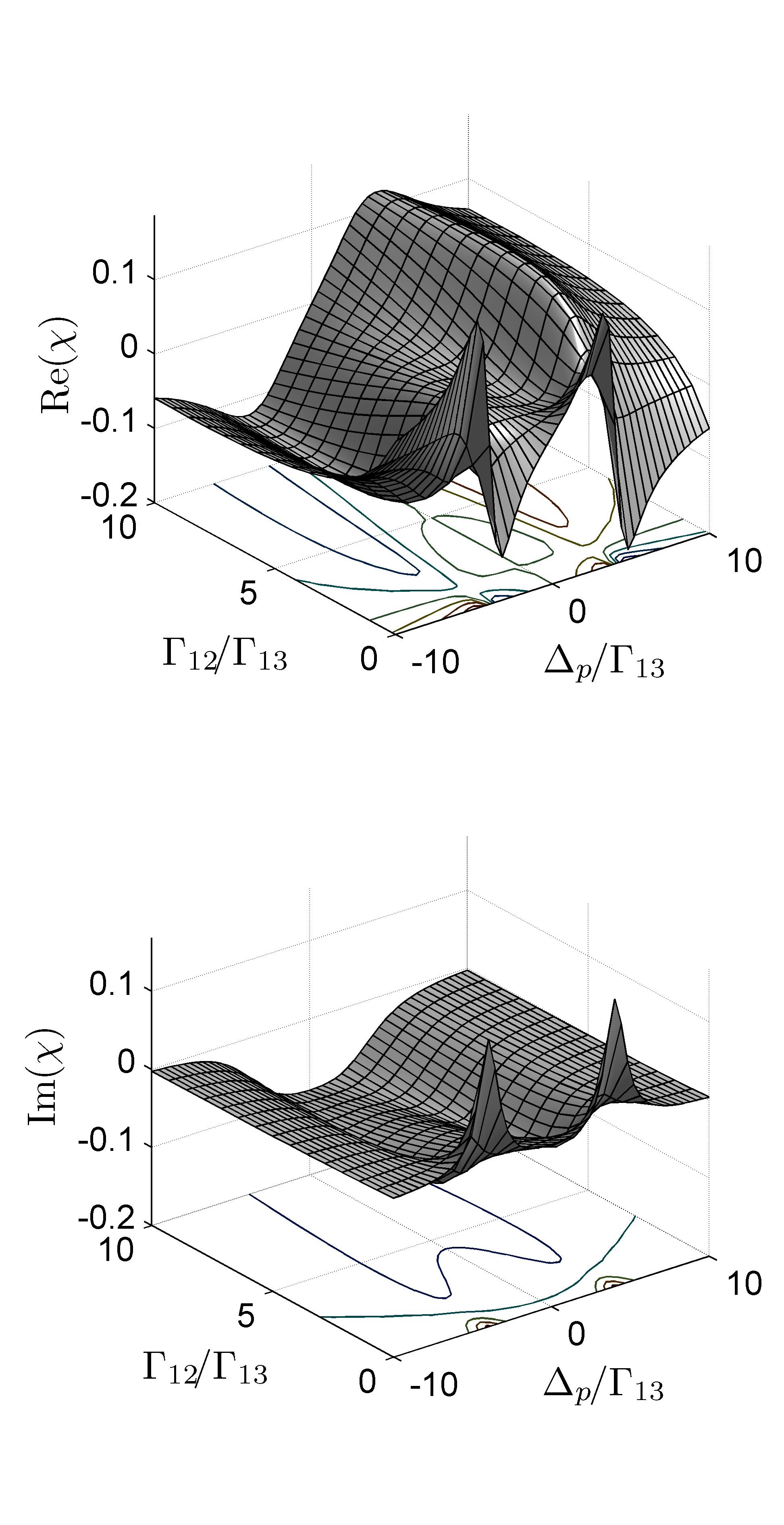} 
\caption{Plot of real (top) and imaginary (bottom) part of the electric susceptibility $\chi_{p}$ (in units of $\Gamma\mu_{13}^2/(V\eps\hbar$)) as a function of the scaled probe detuning $\Delta_{p}/\Gamma_{13}$ and intraband decay rate $\Gamma_{12}/\Gamma_{13}$. \label{fig:intraband}}
\end{center}
\end{figure}
Comparing the two cases (a) and (b) in Fig. \ref{fig:paramInvestigation} , we notice an almost similar slow-down factor for large coupling intensities, but for small intensities the slow-down for case (a) is approximately 2.5 larger than that of (b). This gives insight about the origin of the slow-down effect in the V-scheme: Since the polarization rate in (b) is set very large, the observed slow-down is not due to AT splitting since this requires a coupling Rabi frequency effectively similar to $\gamma_{12}$. Rather, the slow-down originates from \ac{SHB} caused by the coupling field. The hole burning causes the probe to experience a reduced absorption near its resonance which leads to a positive slope of the refractive index. This is confirmed by the variation of the absorption, which is seen to decrease to half its value for increasing coupling intensity. When increasing the Rabi frequency beyond $\gamma_{12}$ ($\sim 10^7 \unit{W/cm^2}$), the absorption decreases further towards  zero due to \ac{ATS}. 
Thus, the observed slow-down effect of the V-scheme for lifetime limited decay, (a), is partly due to AT splitting and partly due to spectral hole burning.  Notably, this is observed in Fig. \ref{fig:slowdownComp}, at low coupling intensities. Here, the slow-down, for decreasing coupling intensity, is seen to decrease more gradually than for the other schemes, due to the \ac{SHB} origin of the slow-down. 

Considering the case of a large intraband population decay rate, (c), we notice that a larger coupling intensity is required at the maximum slow-down. This is a result of a larger value of $\gamma_{12}$ given that is a sum of all three population decay rates. As a result, a larger Rabi frequency is required to resolve the AT splitting. For a large intraband decay rate, the system acts as an amplifier with electrons pumped by the coupling field from $\ket{3}$ to $\ket{2}$ and quickly decaying to $\ket{1}$. Thus, for large coupling intensities ($\sim10^{4}\unit{W/cm^2}$), the probe experiences gain (negative absorption) resulting in a slow-down effect. For the largest intensities, $>10^4\unit{W/cm^2}$, the absorption is seen to return to transparency for increasing coupling intensity as a result of the \ac{ATS}. 
This is illustrated in Fig. \ref{fig:intraband} where the real (left) and imaginary (right) part of the susceptibility is plotted as a function of the probe detuning, for various intraband scattering rates $\Gamma_{12}$.
\begin{figure}
\begin{center}
\includegraphics[width=0.38\textwidth]{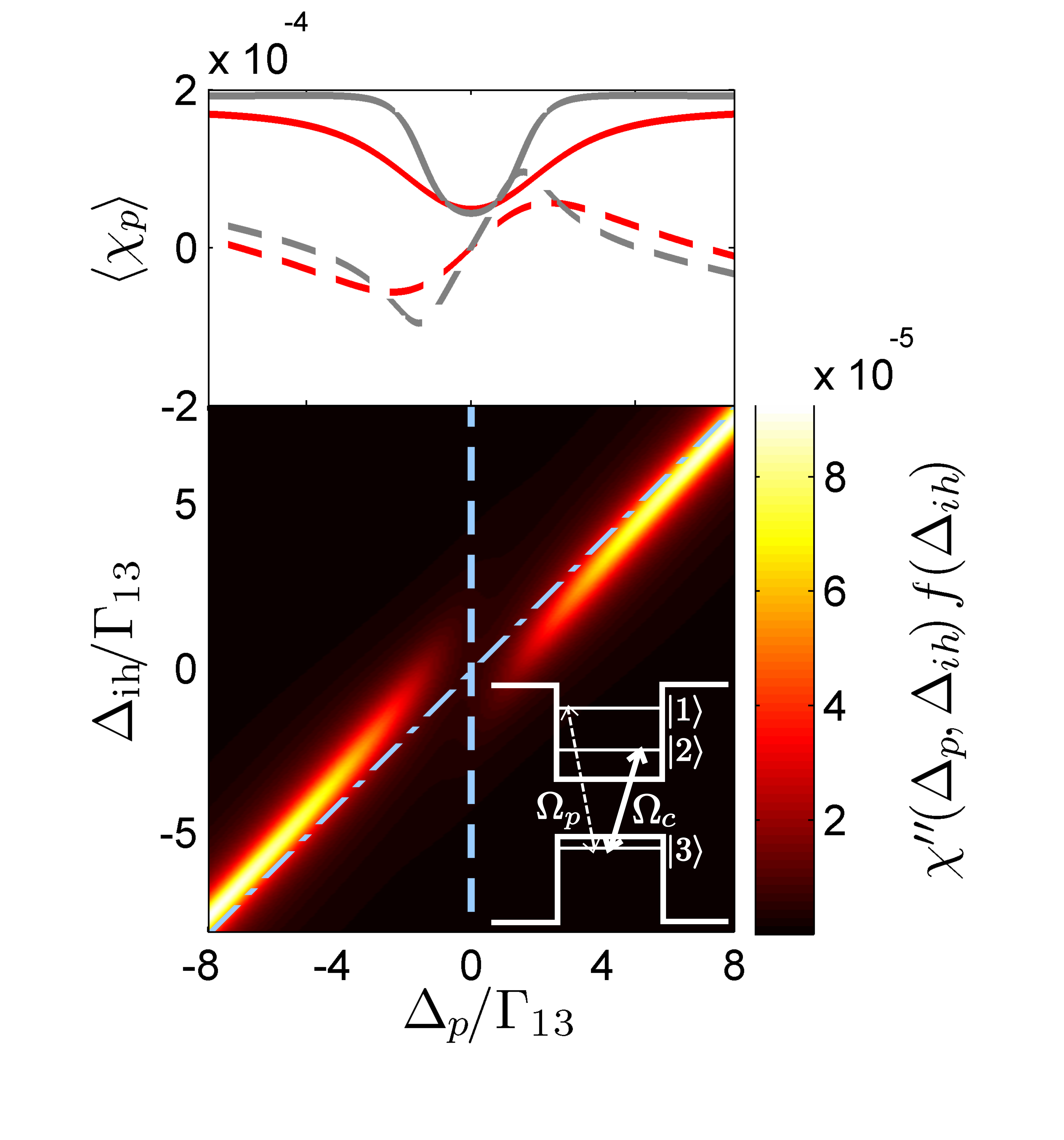} 
\caption{(color online) Calculated electric susceptibility using $\eta=1$. Top: Average real (dashed) and imaginary (solid) part of the electric susceptibility (normalized by $\Gamma\mu_{13}^2/(V\eps\hbar)$) for the alternative V scheme (red) and V-scheme from Fig. \ref{fig:schemes} (grey). Bottom: Associated imaginary part of the electric susceptibility for the alternative V-scheme as a function of scaled probe detuning $\Delta_{p}/\Gamma_{13}$ and spectral shift $\Delta_{ih}/\Delta_{13}$. The inset illustrates the excitation configuration of the alternative scheme. \label{fig:altScheme}.}
\end{center}
\end{figure}
For $\Gamma_{12}=0$, the absorption spectrum shows \ac{ATS} as a result of the strong coupling field. However, as $\Gamma_{12}$ is gradually increased, the absorption becomes negative (i.e. amplification). Considering the real part of $\chi_{p}$ (left), it is seen that the slope remains positive, giving a reduction of the group velocity even for the highest values of $\Gamma_{23}$. It is interesting to note though, that the two effects support each other. Thus, for intermediate values of $\Gamma_{12}$ an effective slow-down may be achieved, while minimizing the absorption due to the small fraction of carriers being pumped to the probe transition.

\subsection{Alternative V schemes}
The V-scheme may also be considered in an alternative configuration where the probe and coupling fields have interchanged transitions as illustrated in the inset of Fig. \ref{fig:altScheme}. We shall denote this the alternative V-scheme. However, there are several  reasons for choosing the standard V scheme illustrated in Fig. \ref{fig:schemes} rather than the alternative configuration. Firstly, the expected weak dipole moment of the "forbidden" transition would result in a small slow-down of the probe, since it scales with the dipole moment square. \cite{Fleischhauer2005} Also, a finite intraband population decay rate would lead to an increased probe absorption as well as a weakened slow-down. More importantly, however, is the associated value of $\kappa$. For the alternative V-scheme we have $\kappa\approx0.43$. Thus, real solutions to \eqref{eq:rescond} exist, i.e. the slope $(1-\kappa)^{-1}$ is positive (see equation \eqref{eq:limres}) leading to an increased absorption of the probe field. This is illustrated in Fig. \ref{fig:altScheme} (bottom) where the probe absorption is plotted as a function of probe detuning $\Delta_p$ and spectral shift $\Delta_{ih}$. The two dressed absorption resonances are seen to overlap the region with $\Delta_p=0$,  and the average spectral dip of the probe absorption near resonance thus becomes less pronounced, resulting in a lower slow-down factor. This is clearly seen in Fig. \ref{fig:altScheme} (top) where the absorption integrated over $\Delta_{ih}$ is plotted as a function of $\Delta_p$ for the two types of V schemes. This clearly demonstrates, that it is not only the \ac{SHB} slow-down effect that makes the V-scheme superior to the other schemes, but also the value of $\kappa$ is important.\\

\subsection{Fine structure splitting schemes}
\begin{figure}
\begin{center}
\includegraphics[width=0.45\textwidth]{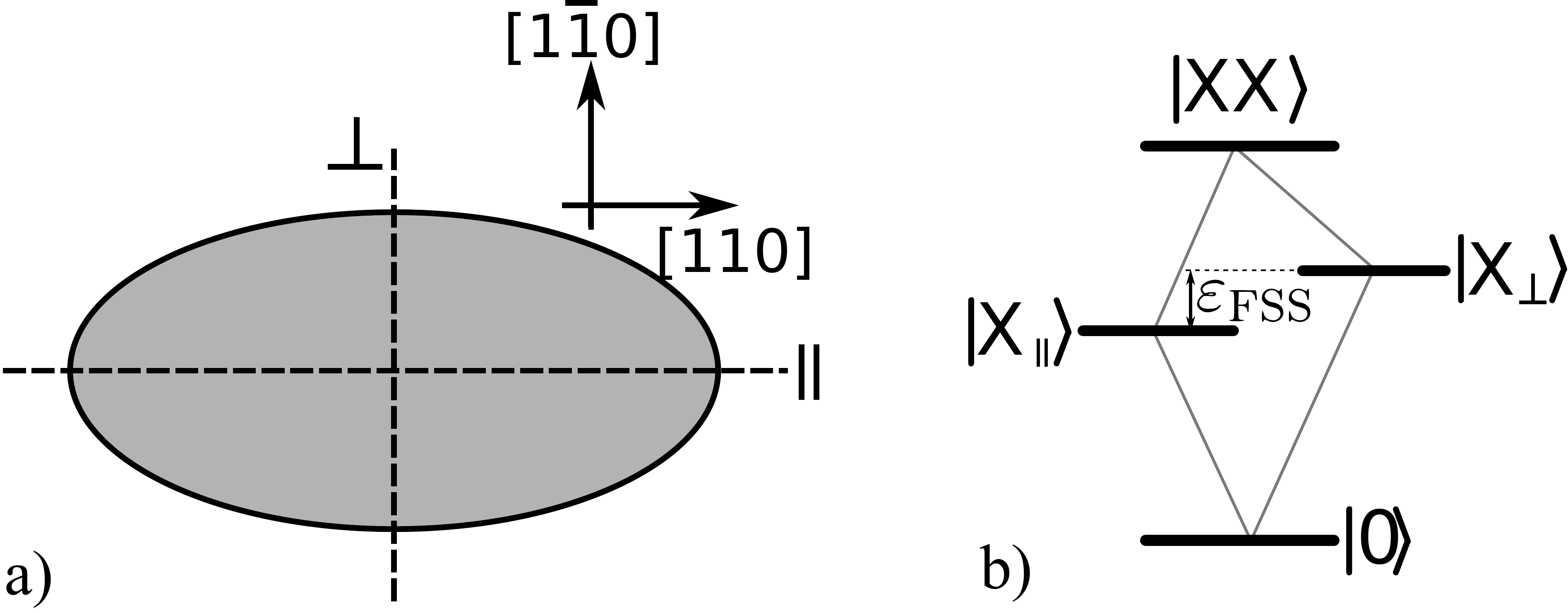} 
\caption{a) Asymmetric QD potential with symmetry axis (dashed lines) along the crystal planes $[110]$ and $[1\bar{1}0]$. b) Fine structure splitting based schemes. The two exciton states $\ket{X_{\parallel}}$ and $\ket{X_{\perp}}$  are coupled with linearly polarized light parallel to the symmetry axis. \label{fig:FSSschemes}}
\end{center}
\end{figure}

As mentioned, \ac{ATS} in \acp{QD} has previously been demonstrated by exploiting the \ac{FSS} in \acp{QD}. The \ac{FSS} arises from the electron-hole exchange interaction that couples the electron spin and hole angular momentum creating two \emph{dark} excitonic states with total angular momentum projection $j_z=j_{h,z}+j_{e,z}=\pm2$ and two \emph{bright} states with $j_z=j_{h,z}+j_{e,z}=\pm1$.\cite{Bayer2002,Michler2003,Seguin2005} The electron-hole coupling leads to a splitting between the dark and bright states, and we shall therefore neglect the former. In a QD with rotational symmetry in the plane of QD growth, the two bright states are degenerate. However, typically QDs have elongated potentials along the crystal axis $[110]$ and $[1\bar{1}0]$, either due to the structural formation of material or due to the strain induced piezoelectric fields.\cite{Gammon1996,Bayer2002,Seguin2005}
When  the rotational symmetry of the QD confinement potential is broken, the angular momentum is no longer a good quantum number and the bright eigenstates evolve into 
a symmetric and antisymmetric linear combinations of the $j_z=\pm1$ basis states, with opposite energy shifts $\frac{1}{2}\varepsilon_{\textrm{FSS}}$. \cite{Michler2003} We shall denote these eigenstates $\ket{X_\perp}$ and $\ket{X_\parallel}$ since they are coupled by light that is linearly polarized perpendicular and parallel to the two axis of reflection symmetry of the \ac{QD}, see Fig. \ref{fig:FSSschemes}a. Considering the biexciton state $\ket{XX}$, where two excitons are excited, we note that this state is not split by the exchange interaction since the net angular momentum of both excitons is 0.\cite{Seguin2005} Together with the crystal ground state $\ket{0}$, these states may be optically driven at the transitions that make up either the $\Xi$- V- or $\Lambda$-scheme, by suitable choice of wavelength and polarization of the coupling and probe field  (see Fig. \ref{fig:FSSschemes}). We shall therefore extend our model to these types of schemes.

Similar to equations \eqref{eq:kappa}, relating the spectral shift of the coupling transition to the probe transition we find for the V-scheme:
\begin{align}
\textrm{V:}\quad\Delta_{ih,c}&=\frac{\delta\left(\varepsilon\lo{X}\pm\frac{1}{2}\varepsilon_{\mathrm{FSS}}\right)}{\delta\left(\varepsilon\lo{X}\mp\frac{1}{2}\varepsilon_{\mathrm{FSS}}\right)}\Delta_{ih,p}\\
&=\left(1\pm\frac{\delta\varepsilon_{\mathrm{FSS}}}{\delta\left(\varepsilon\lo{X}\mp\frac{1}{2}\varepsilon_{\mathrm{FSS}} \right)}\right)\Delta_{ih,p}\\
&\approx\left(1\pm\frac{\delta\varepsilon_{\mathrm{FSS}}}{\delta\varepsilon\lo{X}}\right)\Delta_{ih,p},\label{eq:FSSVkappa}
\end{align}
where $\varepsilon_X$ denotes the exciton energy without the \ac{FSS} and in the last equation it was assumed that $\varepsilon_X\gg\varepsilon_{\mathrm{FSS}}$.
The \ac{FSS}, $\varepsilon_{\textrm{FSS}}$, has previously been reported to vary with the size of the quantum dots,\cite{Seguin2005} i.e. the FSS decreases with increasing groundstate transition energy. However, the magnitude compared to the IHB is rather low, e.g. $\varepsilon_{\textrm{FSS}}$ was found to change approximate $0.5\unit{meV}$ for a shift of the ground state transition of  roughly $300\unit{meV}$.\cite{Seguin2005} Thus, the last term in \eqref{eq:FSSVkappa} is neglected leading to a value of $\kappa\approx1$. 

Analogously we find that $\kappa\approx1$ for the $\Lambda$-scheme.

For the $\Xi$-scheme we get:
\begin{align}
\Xi:\quad\Delta_{ih,c}	&\approx\frac{\delta\varepsilon\lo{XX}}{\delta\varepsilon\lo{X}}\Delta_{ih,p}\\
			&=\left(1+\frac{\delta\left(\varepsilon\lo{XX}-\varepsilon\lo{X}\right)}{\delta\varepsilon\lo{X}}\right)\Delta_{ih,p}
\end{align}
Generally, the biexciton energy, $\varepsilon\lo{XX}$, is different from that of the exciton, $\varepsilon\lo{X}$, due to mixture of Coulomb interaction and correlation effects that depends on the quantum dot size, shape and strain.\cite{Rodt2005} Thus, a change of QD size does not in general lead to identical shifts of the exciton and biexciton transition energies, i.e. $\delta\varepsilon\lo{XX}/\delta\varepsilon\lo{X}\neq 0$. In a series of measurements by S. Rodt et al,\cite{Rodt2005} they measured the energy difference $\varepsilon\lo{XX}-\varepsilon\lo{X}$ as a function of $\varepsilon\lo{X}$. Applying a linear fit to these measurement we find $\delta\left(\varepsilon\lo{XX}-\varepsilon\lo{X}\right)/\delta\varepsilon\lo{X}\approx0.05$.

Comparing the values found for $\kappa$ with the conditions in \eqref{eq:kappacond}, it is immediately seen that the $\Xi$-scheme remains unsuitable for achieving slow light since some QDs contribute to the probe absorption. However, for both the $\Lambda$- and V-schemes a spectral region near $\Delta_p=0$ remains transparent since no dressed state is on resonance with the probe. As mentioned previously, a general problem of exploiting a \ac{FSS} scheme is that it requires the coupling field to propagate perpendicular to the plane of growth. With the $\Lambda$-scheme we are furthermore left with the problem of population decay from the exciton state,  $\ket{X}$, to the crystal ground state $\ket{0}$. Thus, a prepulse is required to optically pump a population into $\ket{X}$. As pointed out in Ref. \onlinecite{Phillips2004}, the prepump needs to arrive sufficiently long time prior to applying the probe field to avoid dipole interference induced by the prepump. This time is governed by the dephasing rate of the $\ket{0}-\ket{X}$ transition. On the other hand, the prepump should not arrive earlier than the relaxation time between $\ket{X_\perp}$ and $\ket{X_\parallel}$, to avoid population in both of the exciton states. Thus, a proper calculation of the electric susceptibility requires solving the density matrix and wave equation in the time domain.\cite{Nielsen2007}  However, restricting our attention to the impact of inhomogeneous broadening, we shall for the $\Lambda$-scheme neglect population decay from $\ket{X}$ to $\ket{0}$. 

\begin{figure}
\begin{center}
\includegraphics[width=0.45\textwidth]{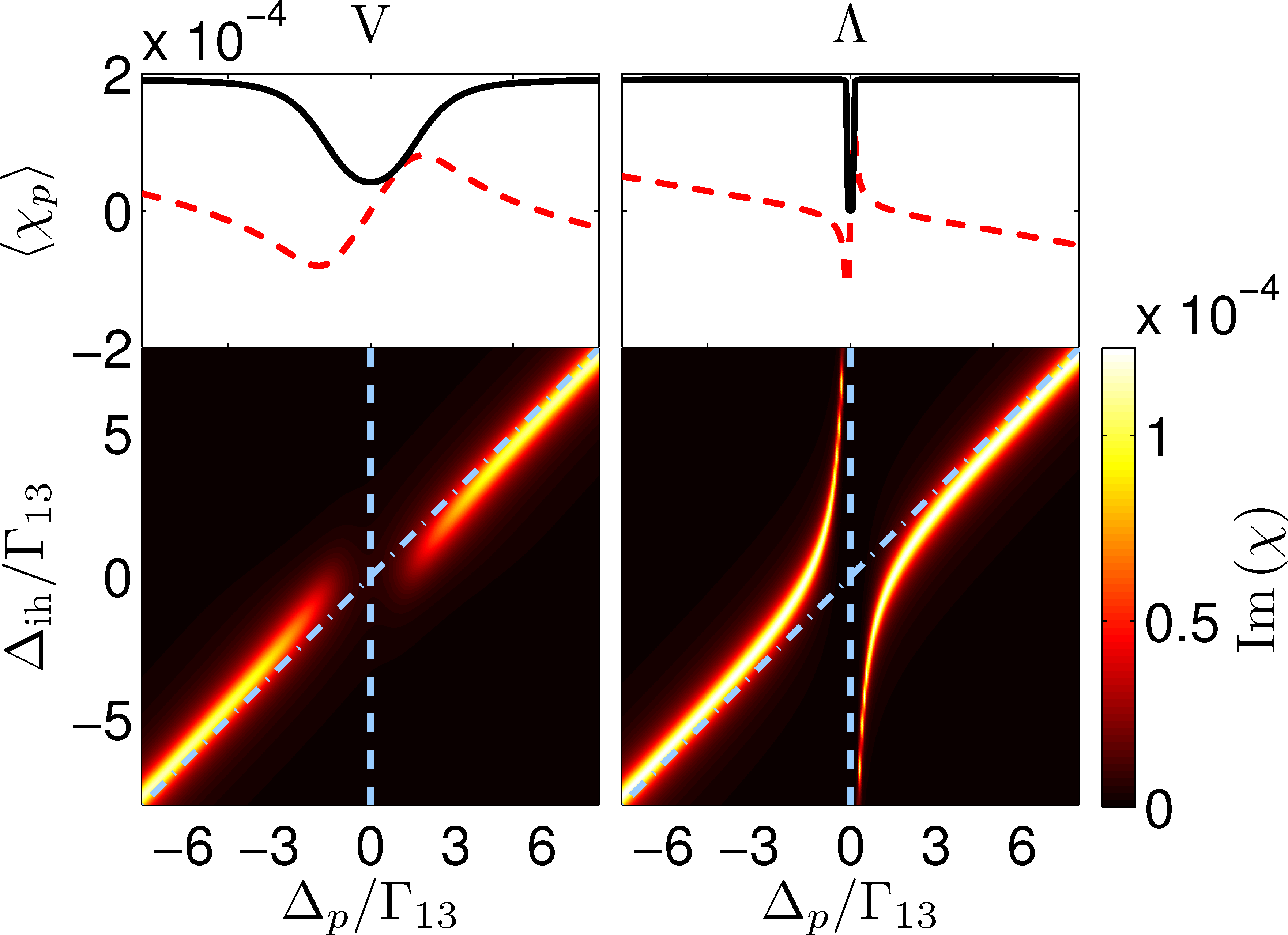} 
\caption{(Color online) Electric susceptibility (in units of $\Gamma\mu_{13}^2/(V\eps\hbar$)) of the V (left) and $\Lambda$ (right) scheme based on the FSS configuration. Bottom:  Imaginary part of the electric susceptibility as a function of the normalized probe detuning, $\Delta_{p}$, and spectral shift $\Delta_{ih}$. The asymptote of the primary and secondary resonance is indicated by a dash-dotted and dashed line, respectively. Top: Corresponding calculated average real (dashed) and imaginary part (solid) part of the electric susceptibility. Key parameters are $\Gamma_p=\Gamma_c$ and $\Gamma_{uc}=0$, where $\Gamma_p$, $\Gamma_c$ and $\Gamma_{uc}$ denote the population decay rate of the probe, coupling and uncoupled transition, respectively. Dephasing rates were assumed lifetime limited.\label{fig:FSSLambdaAndV}}
\end{center}
\end{figure}

The calculated imaginary part of the electric susceptibility is plotted in Fig. \ref{fig:FSSLambdaAndV} for the V-scheme (left) and $\Lambda$-scheme (right) as a function of the probe detuning and spectral shift. 
In both cases we find $\kappa=1$, but the secondary resonance remains absent for the V-scheme due to carrier excitations by the pump, as discussed in section \ref{sec:schemeComp}. When integrating over all spectral shifts, this leads to  significantly different average susceptibilities for the two schemes (Fig. \ref{fig:FSSLambdaAndV} top). Where the V-scheme looks similar to the calculation based on the conduction band states in section \ref{sec:schemeComp}, the $\Lambda$-scheme shows a very sharp and narrow dip in the absorption and consequently a steep slope of the real part of the susceptibility. As a result, the FSS based $\Lambda$-scheme could potentially be used to realize very low group velocities.

\section{Propagation effects}
Since the V-scheme turns out to be robust against \ac{IHB}, the use of a waveguide structure where the electric fields interacts with an ensemble of QDs, is possible. The purpose is two fold: the confinement of the electric fields serves to decrease the nessesary pump power, and furthermore the waveguide provides an efficient way of increasing the interaction length, thereby increasing the potential signal delay. However, as opposed to the two other schemes, the coupling field is subject to carrier absorption. An upper limit on the achievable delay is therefore expected. I.e., for a long waveguide the coupling field is initially very strong and generates a large \ac{ATS} in the first part of the waveguide. The probe absorption in this section is very low and so is the slow-down effect. At the end of the device, on the other hand, the coupling field is absorbed such that the limited \ac{ATS} causes a high degree of absorption, while the slow-down effect on the probe has vanished. Since this problem is avoided with the $\Xi$- and $\Lambda$-scheme, a comparison of the delay and transmission of the three schemes is reasonable.

The propagation effects of both the coupling and probe fields were analyzed using a one-dimensional propagation equation, assuming a single transversal mode. 

The propagation of the probe field was calculated from the wave equation in the slowly varying envelope approximation.\cite{Jackson1998} In the frequency domain the wave equation for the probe field is given as:\cite{Agrawal1989}
\begin{equation}
\frac{\partial\Ef(z,\Omega)}{\partial z}=\ii\left(\frac{\Omega}{u_g}+\frac{\omega_{0}}{2n_{bg} c}\langle\chi_{p}(z)\rangle\right)\Ef (z,\Omega)\label{eq:svea_freq},
\end{equation}
where $z$ is the propagation distance, $\Ef$ is the slowly varying electric field amplitude, $u_g$ is the group velocity of the background material and $\Omega=\omega-\omega_p$. We explicitly indicate the $z$-dependence of $\chi_p$ due to the absorption of the coupling field in the V scheme. Equation \eqref{eq:svea_freq} is formally solved to give:
\begin{multline}
\Ef(z,\Omega)=\Ef(0,\Omega)\cdot\\
  \exp\left[ \ii z\left( \frac{\Omega}{u_g}+ \frac{1}{z}\frac{\Omega+\omega_p}{2n_{bg} c}\int\limits_0^z\langle\chi_{p}(\tilde{z},\omega)\rangle\dif\tilde{z} \right) \right]\label{eq:Esol}.
\end{multline}
Noting that \eqref{eq:Esol} takes the form $\Ef(z)=\Ef(0)\exp(\ii \bar{k})$, where $\bar{k}$ is the mean propagation wavevektor over the propagation length $z$, we can extract the average group index $\langle n_g\rangle$ and absorption coefficient $\langle\alpha\rangle$ using the relations $n_{g}/c=\partial \RE(k)/\partial\omega |_{\omega_{p}}$ and $\alpha=\IM(k)$. Denoting the real and imaginary part of $\chi$ as $\chi\rq{}$ and $\chi\rq{}\rq{}$, respectively, we find the average group index as:
\begin{multline}
\langle n_g(\omega)\rangle= \frac{c}{u_g}+\\
\frac{1}{z}\frac{1}{2n_0 }\int_0^z\left(\langle\chi_{qd}'(\tilde{z},\omega)\rangle+ \omega\frac{\partial\langle\chi '_{qd}(\tilde{z},\omega)\rangle}{\partial\omega}\right)\dif\tilde{z}\label{eq:averagegroupindex},
\end{multline}
while the average absorption coefficient is found as:
\begin{equation}
\langle\alpha\rangle=\frac{1}{z}\frac{\omega_p}{2n_{bg} c}\int\limits_0^z\langle\chi_{p}''(\tilde{z},\omega)\rangle\dif\tilde{z}.\label{eq:averageabs}
\end{equation}
Notice that equation \eqref{eq:averagegroupindex} and \eqref{eq:averageabs} describe the average group index and absorption, respectively, over the propagation length $z$. Using \eqref{eq:averagegroupindex} the aquired probe delay is calculated as
\begin{equation}
\Delta t=\frac{z}{c} \left(\langle n_g(z)\rangle-n_{bg}\right)
\end{equation}
For ease of calculations, the propagation of the coupling field, in the V-scheme, was decoupled from that of the probe field by approximating the pump transition to be a pure 2-level problem, with two decay channels: One directly from $\ket{2}$ to $\ket{3}$ and another via the state $\ket{1}$, where the population decay rate for the latter channel is given by $\left(\Gamma_{12}^{-1}+\Gamma_{13}^{-1}\right)^{-1}$. This approximation is valid for a weak probe  and $\Gamma_{12}\ll\Gamma_{23}$, meaning that the population density of $\ket{1}$ is vanishing. 

\begin{figure}
\begin{center}
\includegraphics[width=0.45\textwidth]{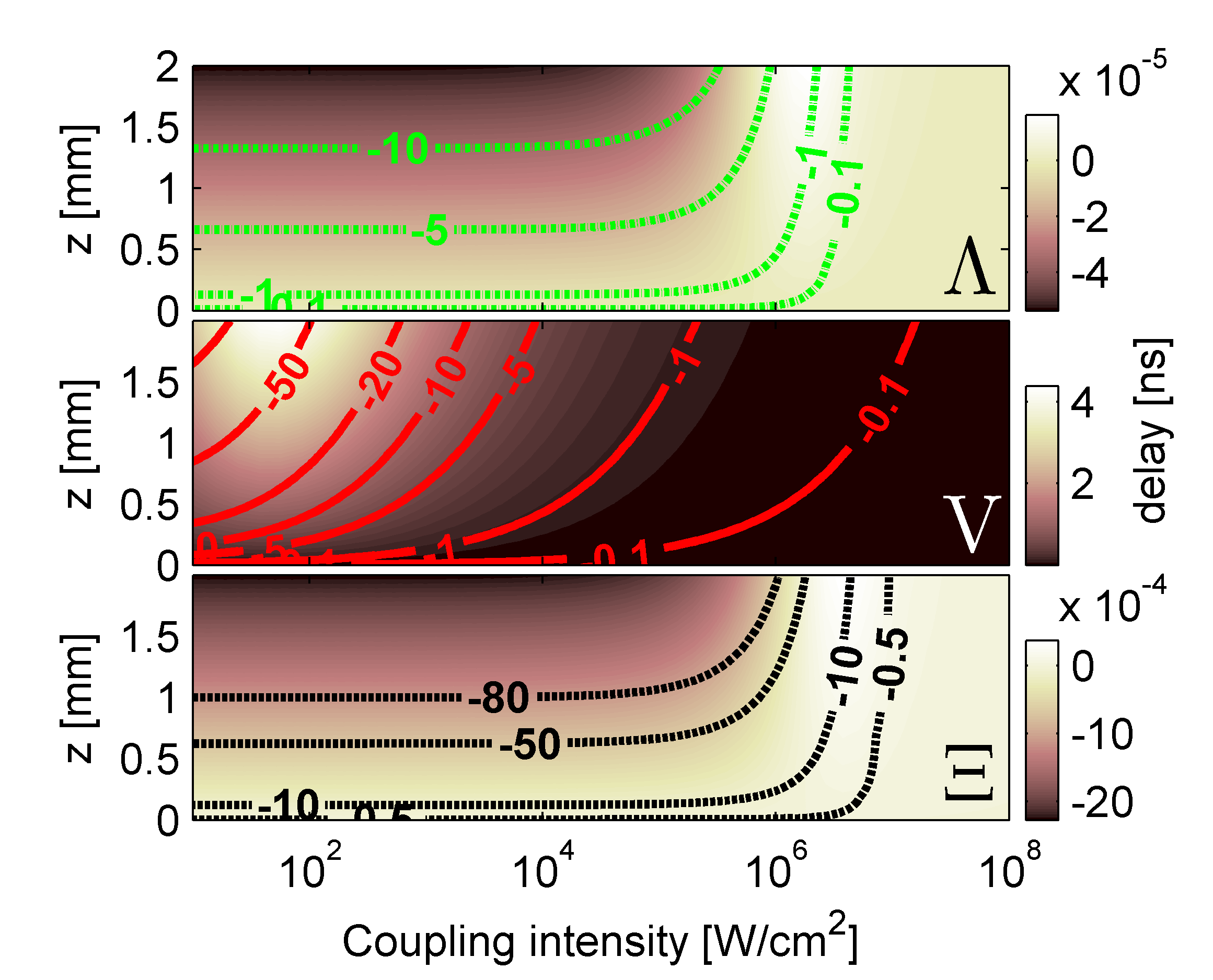} 
\caption{(Color online) Calculated delay (surface plot) and transmission (isocurves) as a function of injected coupling intensity and propagation length $z$ for the $\Lambda$- (top), V- (middle) and $\Xi$-scheme (bottom). The transmission curves are labelled with the corresponding transmission in dB. Note from the colorbar the different scales of the delay. Decay and material parameters are chosen as in table \ref{tab:params}.\label{fig: propagation}}
\end{center}
\end{figure}

In Fig. \ref{fig: propagation} the delay and transmission are plotted as a function of propagation length $z$ and injected coupling intensity for all three schemes $\Xi$, V and $\Lambda$.
The surface plot shows the delay in units of ns, while the lines on top show iso-curves of the calculated transmission in dB.
Noting the different scalings of the associated colors bar, it is seen that for the $\Xi$ and $\Lambda$ scheme, hardly any positive delay is possible as a result of the large \ac{IHB}. For the V scheme, however, a considerable delay, on the order of nanoseconds, is obtained. Furthermore, it is seen that the required coupling intensity for maximum delay, is considerably larger for both the $\Xi$ and $\Lambda$ scheme compared to the V scheme. In fact, such high coupling intensities inevitably requires a strong field confinement such as in a waveguide. However, for these schemes, the coupling transition is typically around $10\unit{\mu m}$ whereas the probe transition is roughly $1\unit{\mu m}$. Fabricating a waveguide for both wavelengths is a challenging task.

Considering the calculated transmission coefficients, the $\Lambda$-scheme shows a considerably larger transmission compared to the other two schemes as a result of the small dipole moment of the probe transition. However, this also affects the achievable delay, being very small. 

\begin{figure}
\begin{center}
\includegraphics[width=0.45\textwidth]{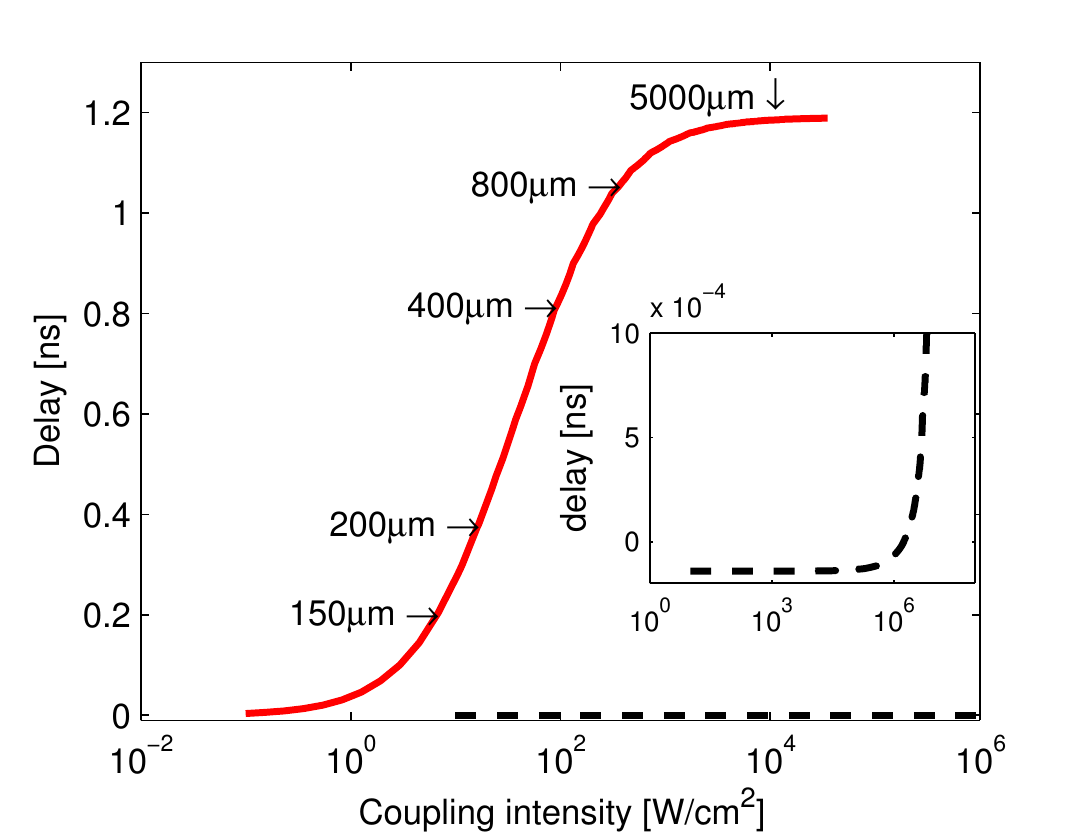} 
\caption{Delay as a function of the injected coupling intensity at a fixed transmission of $-10\unit{dB}$ for the V- (solid) and $\Xi$-scheme (dashed). Corresponding propagation lengths are indicated by arrows for the V- scheme. Inset shows a zoom-in of the calculated delay for the $\Xi$-scheme. Decay and material parameters are chosen as in table \ref{tab:params}.}
\label{fig: delayFixedTransm}
\end{center}
\end{figure}

Inevitably what sets the limit of the achievable delay, is the tolerated absorption of the probe signal. 
A plot of the delay as a function of coupling intensity along the -10 dB transmission contour line is seen in Fig. \ref{fig: delayFixedTransm} comparing the V and $\Xi$ schemes. 
From the inset it is seen that the delay of the $\Xi$ scheme continues to grow for increasing coupling intensity. This is due to the fact that the coupling field does not experience any absorption. Thus, increasing the coupling intensity results in a decreased delay that is compensated by increasing the propagation length. However, group velocity dispersion and waveguide losses, that are not included in the model, set an upper limit on the device length.\cite{Tidstrom2007,Shi2009} As opposed to the $\Xi$- and $\Lambda$-scheme the delay obtained using the V-scheme is seen to approach an upper limit. Thus, the achieved delay is almost constant for intensities above 500 W/cm$^2$ only changing by less than 6 \%. Considering the device length, indicated by arrows, this implies that very little is gained by making the device longer than  $\sim1\unit{mm}$.
Furthermore, it was found that this upper limit roughly scales linearly with the tolerated transmission (in dB). For a Rabi frequency larger than the dephasing rate we have the approximate relations $\Delta t\propto z n_g(z)\propto z \Omega^{-1}$. The AT splitting and therefore also the absorption $\alpha$ is governed by the Rabi frequency of the coupling field. For increasing Rabi frequency, the absorption converges towards zero. Since the maximum delay was found to scale with the logarithm of the tolerated transmission, this implies that $\langle\alpha\rangle\propto\Omega^{-1}$. Thus, for a given tolerated transmission, the field strength of the coupling field should have a magnitude where the average absorption coefficient is roughly inversely proportional to the Rabi frequency in order to achieve the largest delay.

The above calculations are drastically simplified by only considering propagation of monochromatic waves. As such, the calculated delay serves as an upper limit, since group velocity dispersion as well as absorption dispersion may restrict the device length. \cite{Tidstrom2007,Shi2009}

\section{Conclusion}
Slowdown effects in an inhomogeneously broadened quantum dot medium were analyzed for an ensemble of three level QDs strongly driven by a coupling field and probed by a weak field. 

Using a simple but general model for the quantum dot energy levels, the slow-down and absorption of three generic schemes, $\Xi$, V and $\Lambda$, were compared. When neglecting \acl{IHB}, the $\Xi$-scheme seemed advantageous both in terms of slow-down and absorption compared to the other schemes. However, when including \acl{IHB} all slow-down effects vanished for the $\Xi$- and $\Lambda$-scheme, while the V-scheme maintained a slow-down of more than $10^2$. 

Comparing the dressed state eigenenergies it was shown that no quantum dot in the inhomogeneously broadened ensemble would have a dressed state on resonance with the probe in the V-scheme. Thus, as opposed to the other schemes, a transparency region remains near zero detuning of the probe that results in reduction of the group velocity. By analyzing the V-scheme for different intraband relaxation rates, it was further shown that origin of the slow-down effect was an admixture of the \acl{ATS} and spectral hole burning caused by the coupling field. 

Finally, based on field propagation calculations the transmission and signal delay of the three schemes were compared. Despite the finite absorption of the coupling field, the V-scheme showed significantly larger achievable delays. For a fixed transmission of -10 dB the delay was seen to converge towards a constant value ($\sim 1\unit{ns}$) for an increasing coupling intensity. As a result, the delay was almost constant for propagation lengths beyond $1\unit{mm}$.

The theory and simulations presented in this paper indicate the possibility, by proper tailoring of the level scheme and level lifetime, of experimentally achieving optically controlled slow light in an inhomogeneously broadened quantum dot medium using a waveguide structure for confinement of both probe and coupling field.


\begin{acknowledgments}
This work was supported by the Danish Research Councils in the framework of QUEST as well as the European Commision via the FP7 FET project GOSPEL.
\end{acknowledgments}

\acrodef{EIT}{electro magnetically induced transparency}
\acrodef{IHB}{inhomogeneous broadening}
\acrodef{QD}{quantum dot}
\acrodef{ATS}{Autler-Townes splitting}
\acrodef{FWHM}{full width at half maximum}
\acrodef{SHB}{spectral hole burning}
\acrodef{FSS}{fine structure splitting}

%

\end{document}